\documentclass [ashowpacs,preprintnumbers,amsmath,amssymb,twocolumn]{revtex4}

\usepackage{hyperref}

\usepackage{graphicx}

\usepackage{dcolumn}
\usepackage{amssymb}
\newcommand{\be}{\begin{eqnarray}}
\newcommand{\ee}{\end{eqnarray}}

\newcommand{\np}[1]{Nucl. Phys. {\bf #1}}

\begin{document}

\title {The interplay of nuclear and Coulomb effects in proton breakup from exotic nuclei}

\author{Ravinder Kumar$^{a,b}$ and  Angela  Bonaccorso$^{a}$ \\ \small
  $^{a}$INFN, Sez. di Pisa, Largo Pontecorvo 3, 56127
Pisa, Italy.\\ \small    $^{b}$Dep. of Physics, Deenbandhu Chhoturam Univ. of Science and  Technology,\\
Murthal, Sonepat, Haryana 131039, India.}

\begin{abstract}
This paper  gives new insight  to the study of dynamical effects in proton breakup as compared to neutron breakup from a weakly bound state in an exotic nucleus.   Following our recent work \cite{ravme} there has been some discussion in the literature \cite{bras,morokuc}, thus in order to clarify and asses quantitatively which mechanism would dominate measured observables, we  study here several reaction mechanisms  separately but also their total including interference.  These mechanisms are:  the recoil effect of the core-target Coulomb potential which we distinguish from the direct proton-target Coulomb potential and nuclear breakup, which consists of  stripping  and diffraction. Direct Coulomb breakup  typically gives cross sections about an order of magnitude larger than the recoil term and the amount of nuclear diffraction vs. Coulomb depends on the target.   Thus for each mechanism the absolute values of breakup  cross sections and parallel momentum distributions for $^{8}$B and $^{17}$F projectiles calculated on a light and a heavy target in a range of intermediate incident energies (40-80A.MeV) are presented. Furthermore we study  in detail the interference among the two Coulomb effects and nuclear diffraction.  The calculation of the direct and recoil Coulomb effects separately and of their interference is the new and most relevant aspect of this paper.  \end{abstract}

\maketitle

{\bf Pacs} {21.10.Jx, 24.10.-i, 25.60.Gc, 27.30.+t}

In a previous publication \cite{ravme} we  studied
dynamical effects in proton breakup  from a weakly bound state in an exotic nucleus on a heavy target. We used a semiclassical method that treats the full  Coulomb and nuclear interactions to all orders \cite{nois,nois2}. The dynamics of proton nuclear and Coulomb breakup was compared to that of an {\it equivalent} neutron of larger binding energy in order to elucidate the differences with the well understood neutron  breakup mechanism. We found that with respect to nuclear breakup a proton behaves exactly as a neutron of larger binding energy. The extra "effective energy" is due to the combined core-target Coulomb barrier (cf. Fig.2 of Ref.\cite{ang04}). In Coulomb breakup we distinguished in Ref.\cite{ravme} the effect of the core-target Coulomb potential (called recoil effect), with respect to which the proton behaves again as a more bound neutron, from the (direct) proton-target Coulomb potential effect. The latter gave cross sections about an order of magnitude larger than the recoil term. However the much debated \cite{ang04} question of the relative magnitude of  nuclear and  Coulomb breakup, was not assessed from a quantitative point of view. This question as been raised again \cite{bras} in relation to a study, via the  Continuum Discretized Coupled Channel method (CDCC), of the effect of breakup on elastic scattering. We will show in this paper how reaction theory can presently answer such a question.
\begin{figure}[h]
\includegraphics [height=.28\textheight]{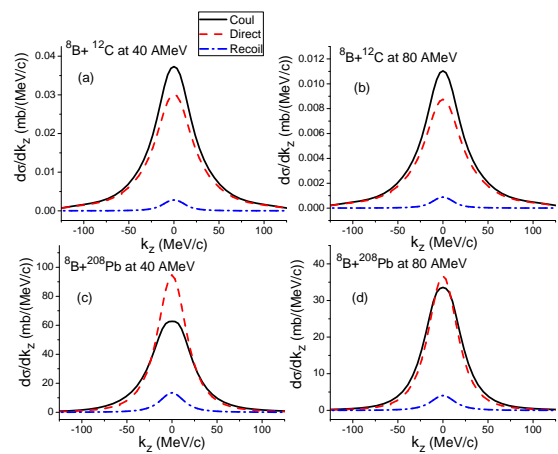}
\caption{(Colour online) Parallel momentum distributions due to the Coulomb recoil  and direct terms  from  $^8$B on $^{12}$C and $^{208}$Pb as indicated and their combined effect including interference.}\label{f1}
\end{figure}

 \begin{figure}[h]
\includegraphics [height=.28\textheight]{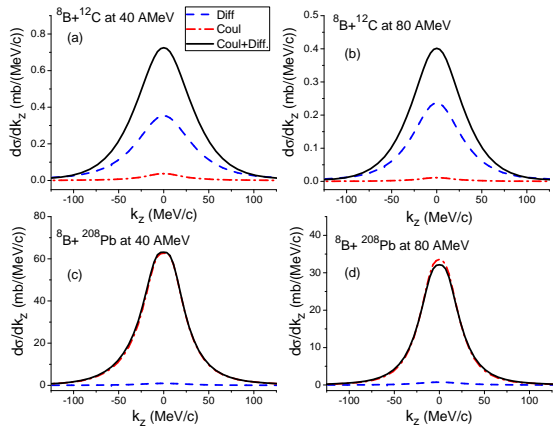}
\caption{(Colour online) Parallel momentum distributions due to diffraction and  Coulomb breakup  from   $^{8}$B on $^{12}$C and $^{208}$Pb as indicated and their combined effect including interference.}\label{f2}
\end{figure}

In fact,  in another recent paper \cite{morokuc} in which CDCC has also been used, the authors have shown that our predictions \cite{ravme} for the proton angular distribution after Coulomb breakup were correct and thus they have validated our interpretation. Such comparison is interesting for two reasons. The first is that CDCC is easier to apply at low energy because the convergency is faster while a semiclassical method like ours works well in the medium-high energy domain. The second is that in our method the interpretation of the results is straightforward because the relative motion between projectile and target is introduced via a semiclassical trajectory and thus the Coulomb and nuclear potential that the breakup particle feels can be treated exactly and approximations can be checked, without disturbing the relative motion treatment.  Thus the two methods could be complementary in their applications. 
 
\begin{table}[h]
\caption{Barrier radii, initial binding energies and effective energy parameters for a $^{208}$Pb target.}
\vskip.3in \begin{center}
\begin{tabular}{lccccccc}
\hline\
                     & $^{8}{\rm B}$&J$^\pi$& $^{17}{\rm F}$&J$^\pi$\\ \hline
R$_{i}$(fm) & 6.0 && 6.5&\\
$\varepsilon_i$(MeV) & -0.14 &1p$_{3/2}$    & -0.6 &1d$_{5/2}$ \\
$-\Delta$(MeV)&-0.4&&-1.2&&\\
$\tilde{\varepsilon}_i$(MeV) & -0.54 &1p$_{3/2}$    & -1.8 &1d$_{5/2}$ \\
\hline
\end{tabular}\end{center}
\end{table}
 We then proceed 
in this paper to present  the  calculated absolute values of the cross sections due to  the nuclear and Coulomb breakup (recoil and direct) separately and then show how much the interference effects modify the simple sum of the cross sections. This is very important in view of spectroscopic studies of proton vs. neutron rich nuclei 
\cite{[1]}-\cite{suz1} and also for the applications in nuclear astrophysics since Coulomb breakup is considered the inverse process of the (p,$\gamma$) capture \cite{[1]}. Results from breakup  on a light,  and a heavy target will be discussed in a range of incident energies from 40 to 80AMeV. The details of the theory can be found in  \cite{nois,nois2}.

Table I shows the bound state parameters used in the calculations. Although the calculations are done here with the exact proton wave function we give also the effective binding energies discussed in Ref.\cite{ravme} to help the reader understanding the difference with the neutron breakup. Spectroscopic factors for the initial states are taken equal to one. For both projectiles only breakup from the valence state is considered. All other parameters used in the calculations are the same as in our previous papers \cite{ravme,nois2}.

Table II contains the absolute values of the cross sections  for the  one proton breakup from $^{8}$B and $^{17}$F on $^{12}$C and $^{208}$Pb at 40, 60 and 80 A.MeV. The cross sections due to the stripping and diffraction mechanisms of the nuclear breakup and the direct and recoil terms of the Coulomb breakup are shown separately. We give also the total Coulomb cross sections which contain the interference effects of direct and recoil terms. Furthermore the total elastic breakup  (diffraction plus Coulomb) cross sections are given. They contain all  interference effects between the three possible mechanisms (nuclear, direct Coulomb, recoil Coulomb) following which  the proton would be measurable in coincidence with the core. We remind the reader that nuclear stripping instead  is the mechanism in which the nucleon is emitted by the projectile and undergoes a final state inelastic scattering with the target. It is thus considered absorbed by the target, in the sense of the optical model absorption and its energy degraded such that it would not be detected in coincidence with the core of origin. Such mechanism cannot interfere with diffraction nor with Coulomb breakup. Stripping is larger than diffraction, as first noticed in Ref.\cite{ang91}.

Parallel momentum distributions due to the Coulomb recoil  and  Coulomb direct terms  from  $^8$B and  $^{17}$F  and their combined effect including interference are shown in Fig.1 and 3 respectively, while Fig.2 and 4 show parallel momentum distributions due to nuclear and Coulomb breakup  from  the same projectiles  and their total effect including interference. Notice that in Fig.3 some asymmetries appear due to the interference  of the direct and recoil Coulomb effects.

\begin{figure}[h]
\includegraphics [height=.28\textheight]{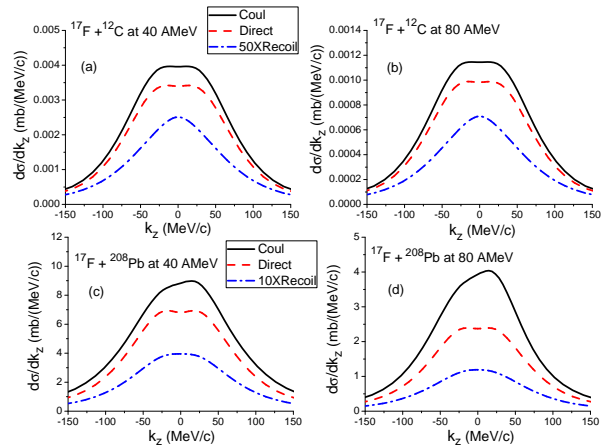}
\caption{(Colour online) Parallel momentum distributions due to the Coulomb recoil  and direct terms  from  $^{17}$F on $^{12}$C and $^{208}$Pb as indicated and their combined effect including interference.}\label{f4}
\end{figure}


\begin{figure}[h]
\includegraphics [height=.28\textheight]{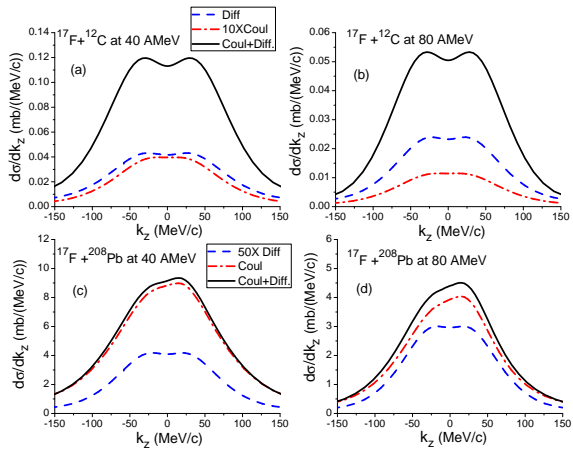}
\caption{(Colour online) Parallel momentum distributions due to diffraction and  Coulomb breakup  from   $^{17}$F on $^{12}$C and $^{208}$Pb as indicated and their combined effect including interference.}\label{f5}
\end{figure}

In the case of the $^8$B projectile at 40AMeV incident energy on the $^{208}$Pb target both the cross section  values in Table II and  figure  1c  show that the direct and recoil Coulomb term interfere destructively and total Coulomb is almost exactly the difference of the two. Increasing the incident energy,  the two Coulomb effects show very small  interference and the total is very close to the sum of the two in the total cross section (cf.Table II) while in the momentum distributions  shown in Fig.1d  at the very small parallel momentum  values it is given by the difference of the two with the recoil term just contributing more. The interference between diffraction and Coulomb is also very small and it is destructive or constructive depending on the incident energy on the heavy target, Figs 2c and 2d.  As expected, on the light  $^{12}$C target   the recoil effect is really negligible and the Coulomb breakup is mainly due to the direct term at all incident energies, Figs.1a and 1b. Thus the interference is small and always constructive.  Diffraction cross sections on the other hand have much higher values  than Coulomb breakup cross sections for the light target. The interference is so strong at low energy that it almost doubles the simple sum of diffraction and Coulomb breakup, Figs 2a and 2b. This effect is very interesting and it shows that including the Coulomb breakup the cross section can increase a lot but not because the Coulomb itself is large, but because of the interference. 

In the case of the $^{17}$F projectile the effects are similar but the interference, in the cases shown here, is always constructive   both between direct and recoil Coulomb as well as between Coulomb and diffraction as can been seen from Figs. 3 and 4.

On the other hand looking at Table II one sees also that for both projectiles the total nuclear breakup cross sections are always of the same order of magnitude than recoil Coulomb breakup on a heavy target but much smaller than the direct Coulomb and the total Coulomb cross sections. Thus we confirm what has already suggested by other authors \cite{[12,esb02,hus}, on why in the past  calculated nuclear breakup of a proton has been found comparable or even larger than the Coulomb breakup. The misinterpretation was simply due to a underestimate of the direct Coulomb breakup due to both the dipole approximation and its treatment to first order and to the fact that  interference effects were overlooked. In particular our present results and interpretation seem to corroborate the CDCC calculations and interpretation of Ref.\cite{hus}, the new aspect of our method being 
the study of the direct and recoil Coulomb effects separately and of their interference. 

In conclusion, in this paper we have presented  results of  calculations for all mechanisms that can produce breakup of a weakly bound proton from an exotic nucleus impinging  on a light and a heavy target. The semiclassical method used allows to treat both the full  nuclear and Coulomb interactions to all orders and all multipolarities. On a light target the total nuclear breakup is always larger than the Coulomb breakup. On the other hand although the Coulomb breakup is very small the interference between diffraction and Coulomb is constructive and such that the total becomes quite large. On a heavy target instead the total nuclear breakup is of the same order of magnitude as the Coulomb recoil effect while the direct Coulomb breakup is one order of magnitude larger. Thus this term dominates not only in the total Coulomb breakup but also in  the total diffraction plus Coulomb term. The quantitative assesment of the direct Coulomb breakup and of its interference with other mechanisms is very important and given here for the first time in the literature.  It is then clear that the breakup mechanism of a proton
is much more complicated than that of a neutron and disentangling various effects is of fundamental importance when interpreting experimental data. Interference effects are somehow impossible to predict without an explicit calculation and as it has been shown above might vary from one observable to the other and  very accurate yet simple to interpret reaction models are  necessary to analyze data and/or to make predictions in order  to plan future experiments. This is particularly  true for applications in nuclear astrophysics where Coulomb breakup is considered the inverse process of the (p,$\gamma$) reaction. Such a concept  will have to be handled with great care in the future. Detailed calculations such as those presented here or made with CDCC, depending on the incident energy, should be performed and correctly  interpreted in order to asses two aspects: i) if and  which  part of the cross section could be considered corresponding to the (p,$\gamma$) reaction cross section; ii) if such separation could be also done by an appropriate experimental procedure and in the data.
\newpage
\begin{table}[h]
\begin{widetext} 
\caption{$ \sigma_{b_{up}}$(mb) for nuclear and Coulomb mechanisms as indicated, for $^{8}$B, 1p$_{3/2}$ initial state, and $^{17}$F, 1d$_{5/2}$ initial state, on $^{12}$C and $^{208}$Pb targets at E$_{inc}$=40, 60, 80MeV.}
\begin{center}
\begin{tabular}{|c|ccc|ccc|}\hline
 Target&&$^{12}$C &&&$^{208}$Pb&\\
 \hline
 E$_{inc}$(A.MeV)&40&60&80&40&60&80\\ 
 \hline\
Projectile   & $^{8}$B ~~~ $^{17}$F&$^{8}$B ~~~ $^{17}$F&$^{8}$B ~~~ $^{17}$F& $^{8}$B ~~~ $^{17}$F&$^{8}$B ~~~ $^{17}$F&$^{8}$B ~~~ $^{17}$F\\ \hline
Stripping & 51.62~~~18.06&41.17~~~13.49&34.79~~~10.93&105.94~~~29.97&88.59~~~23.09&78.16~~~19.29\\
Diffraction &31.72~~~8.19&23.16~~~5.42&18.86~~~4.15&70.42~~~14.08& 58.84~~~10.99&52.39~~~9.36\\
Coulomb recoil &0.10 ~~~0.007 &0.05~~~0.004&0.03~~~0.002&534.18~~~65.98&262.23~~~31.74&159.09~~~19.14\\
Coulomb direct&2.09~~~0.58&1.01~~~0.28&0.61~~~0.17&4562.66~~~1209.35& 2578.76~~~624.61& 1741.04~~~394.54\\
Total Coulomb &2.51~~~0.67 & 1.21~~~0.32& 0.73~~~0.19&4129.47~~~1542.39&2796.84~~~874.40& 1925.34~~~611.52\\
 Coulomb and Diffraction &60.29~~~22.79&39.74~~~13.18& 30.89~~~9.42&4228.56~~~1608.39& 2740.82~~~956.64&1928.03~~~691.09\\
\hline
\end{tabular}\end{center}\end{widetext}
\end{table}

\end{document}